# Token Optimization Strategies for LLM-Based Oracle-to-PostgreSQL Migration

Oleg Grynets
EPAM Systems
McLean, Virginia, USA
oleg_grynets@epam.com

Dmytro Babarytskyi
EPAM Systems
Kharkiv, Ukraine
dmytro_babarytskyi@epam.com

Vasyl Lyashkevych
EPAM Systems
Lviv, Ukraine
vasyl_lyashkevych@epam.com

***Abstract*—Large Language Models (LLMs) are increasingly used for software modernization, code translation, automated refactoring, and database migration. However, LLM-based Oracle-to-PostgreSQL migration remains constrained by high token consumption, long-context degradation, dialect-specific semantic differences, and the risk of semantic drift during query transformation. Direct inclusion of large Oracle SQL/PL-SQL artefacts, schema definitions, procedural logic, and migration instructions into the model context increases cost and may reduce generation quality. This paper investigates token optimization as a constrained transformation problem in LLM-based Oracle-to-PostgreSQL migration. The study formalizes and evaluates twelve token optimization strategies: baseline representation, context pruning, minification, DSL-based semantic compression, metadata augmentation, context refactoring, schema distillation, adaptive routing, AST-based minification, identifier masking, output constraint enforcement, and hybrid optimization. The strategies are evaluated on samples of 10 and 100 Oracle SQL queries using Valid Syntax Rate, Exact Match, Semantic Match, CodeBLEU, and Token Efficiency. The results show that mild context pruning preserves semantic quality almost at baseline level, achieving 89.75% Semantic Match on the 100-query sample compared with 89.80% for the unoptimized baseline. Adaptive routing provides the best practical trade-off, reducing input tokens by 8.72% and output tokens by 5.49% while maintaining 88.40% Semantic Match and increasing Token Efficiency by 6.67%. In contrast, aggressive schema distillation increases Token Efficiency by 132.22% but causes a 44.50 percentage-point decrease in Semantic Match. The findings demonstrate that token optimization cannot be treated as simple prompt shortening; it must be evaluated as a multi-objective migration problem balancing cost, syntactic validity, semantic preservation, and structural fidelity.**

## I. Introduction

Large Language Models (LLMs) have significantly changed software engineering by supporting code generation, code translation, automated refactoring, program repair, documentation generation, test synthesis, and database migration. In enterprise modernization scenarios, LLMs are increasingly considered for transforming legacy database systems, including Oracle SQL/PL-SQL artefacts, into open-source database technologies such as PostgreSQL [1]. This direction is related to broader progress in LLM-supported Text-to-SQL generation, semantic parsing, constrained SQL synthesis, and database-oriented code generation [2], [3].

Oracle-to-PostgreSQL migration is a particularly difficult case for LLM-based transformation because it involves not only syntactic conversion between SQL dialects but also preservation of procedural logic, schema constraints, triggers, stored procedures, packages, cursors, exception handling, built-in functions, storage parameters, and execution semantics. Although LLMs can generate plausible PostgreSQL code, direct query-to-query transformation may hide semantic drift, omit implicit constraints, or produce syntactically valid but functionally non-equivalent output. This problem becomes more complex when large schemas, procedural blocks, comments, migration rules, examples, and target-platform constraints are inserted into the model context.

A common approach to LLM-mediated database migration is direct inclusion of raw Oracle SQL/PL-SQL code and supporting schema artefacts into the prompt. This approach is simple but operationally expensive. It increases input token consumption, raises API cost, and may reduce generation quality when the context becomes too long. Long-context studies show that LLMs may fail to use relevant information when it is buried inside large prompts, a phenomenon known as “lost in the middle” [4]. Prompt compression methods such as LLMLingua and LongLLMLingua demonstrate that prompt length can be reduced, but general-purpose compression does not automatically guarantee semantic preservation in structurally constrained languages such as SQL [5], [6].

The token optimization problem is especially important in specification-driven development (SDD). In SDD, specifications and requirements become the primary artefacts for code generation, while implementation is delegated partly or fully to automated tools or AI agents. A typical SDD workflow includes context acquisition, code parsing, vector search, generation and structuring of LLM responses, agent orchestration, and validation in CI/CD pipelines. This means that the LLM context may contain not only source code, but also specifications, schema fragments, migration rules, validation constraints, tool outputs, and agent instructions. Therefore, token optimization becomes a scientific and engineering problem: how to reduce the context without losing the semantic information required for correct migration.

Existing studies address prompt compression [5], [6], repository-level retrieval [7], Text-to-SQL generation [2], [8], constrained decoding [9], and code representation learning [10]–[15]. However, the specific problem of token optimization for Oracle-to-PostgreSQL migration remains insufficiently studied. Most existing approaches focus either on general prompt compression, natural language reasoning, repository-level code completion, or SQL generation from natural language. They do not systematically evaluate how different context transformation strategies affect SQL dialect migration quality, semantic equivalence, syntactic validity, and token efficiency.

This paper addresses this gap by evaluating twelve token optimization strategies for LLM-based

Oracle-to-PostgreSQL migration. The study treats each strategy as a transformation function applied to the input context or system prompt. The strategies are evaluated using a multidimensional protocol that includes input and output token counts, Valid Syntax Rate (VSR), Exact Match (EM), Semantic Match (SM), CodeBLEU, and Token Efficiency (TE).

The main contributions of this paper are as follows:

- A taxonomy of twelve token optimization strategies for LLM-based Oracle-to-PostgreSQL migration is proposed.
- The token optimization problem is formalized as a constrained transformation task inside specification-driven development.
- An empirical comparison is conducted on 10-query and 100-query Oracle SQL samples.
- The study identifies practical “safe” strategies, such as context pruning and adaptive routing, and high-risk strategies, such as schema distillation and identifier masking.
- A decision-oriented model is proposed for selecting token optimization strategies based on the structural profile of Oracle SQL/PL-SQL artefacts.

## II. Overview of AI-Driven Development Tools and Frameworks

AI-driven development is rapidly moving from isolated code completion toward integrated development ecosystems that combine code editors, CLI agents, IDE extensions, multi-agent orchestration, testing, security validation, project management, and model-context integrations. The analysed tool landscape includes AI code editors, terminal agents, IDE extensions, multi-agent frameworks, code generation systems, testing and security tools, MCP servers, code review tools, knowledge management tools, and specification-driven development platforms.

Table 1 summarizes the general tool groups and their relevance to token optimization in LLM-based migration.

This overview demonstrates that token optimization is no longer limited to prompt engineering. In modern AI-driven development pipelines, tokens are consumed by specifications, code, metadata, tool descriptions, orchestration instructions, retrieved documents, tests, review prompts, and generated outputs. Consequently, the migration problem must be analysed in the context of the entire AI-assisted development workflow.

TABLE I. Overview of AI-Driven Development Tools and Their Relevance to Token Optimization

| Tool group | Representative examples | Main role | Relevance to token optimization |
|---|---|---|---|
| AI code editors and IDEs | Cursor, Windsurf, Kiro, Void, PearAI | Interactive AI-assisted coding and refactoring | Require compact and relevant project context |
| Terminal and CLI agents | Aider, Claude Code, OpenAI Codex CLI, Gemini CLI, Qwen Code | Command-line coding, debugging, migration, and test execution | Token cost grows with repository and schema size |
| IDE extensions | Continue, Cline, Roo Code, GitHub Copilot extensions | Embedded code generation and review | Need local context selection and prompt control |
| Multi-agent orchestration | AutoGen, CrewAI, LangGraph, OpenHands, MetaGPT | Role-based task decomposition and collaborative generation | Multi-agent workflows multiply token consumption |
| SDD tools | fspec, OpenSpec, Spec Kit, spec-driver, MetaSpec, SHOTGUN | Specification-first development and code generation | Specifications become part of the LLM context |
| MCP and integration tools | GitHub MCP, database MCP, Docker MCP, Context7 MCP | Documentation, connect LLMs to tools, repositories, databases | Tool descriptions and retrieved artefacts increase token load |
| Testing and security tools | Promptfoo, PR-Agent, qodo-cover, AgentLint, VibeSec | Prompt testing, code review, validation, and security checks | Validation artefacts require additional context |
| Code search and RAG tools | LlamaIndex, Chroma, Qdrant, repository maps | Retrieve relevant code and documentation | Poor retrieval increases prompt noise and semantic drift |

## III. Token Optimization Within Specification-Driven Development

Specification-driven development changes the role of source code in the software development process. Instead of treating code as the only authoritative artefact, SDD treats requirements, specifications, architecture decisions, data schemas, tests, and validation rules as first-class inputs to automated generation. This has direct consequences for Oracle-to-PostgreSQL migration.

In a traditional migration pipeline, the input may consist mainly of Oracle SQL/PL-SQL code. In an SDD-oriented LLM pipeline, the model may receive:

- business requirements;
- migration specifications;
- Oracle schema definitions;
- PostgreSQL target constraints;
- dialect transformation rules;
- examples of previous migrations;
- validation criteria;
- tool outputs;
- repository context;
- agent role instructions.

Therefore, the token optimization problem becomes a problem of selecting or transforming the minimum sufficient context required for correct migration. Excessive context increases cost and may reduce performance, while insufficient context may lead to semantic drift.

Let $Q_o$ denote the original Oracle SQL/PL-SQL artefact, $S$ denote the specification and context bundle, $f_i$ denote a token optimization strategy, $G$ denote the LLM generation function, $P$ denote the user prompt, $H$ denote the system prompt, $Q_p$ denote the generated PostgreSQL artefact, and

$Q_r$ denote the reference PostgreSQL artefact. Then the migration process can be represented as:

$$Q_p = G(f_i(Q_o, S), H). \quad (1)$$

The optimization objective is not simply to minimize token count. Instead, it is to find a transformation $f_i$ that reduces token cost while preserving syntactic validity and semantic equivalence:

$$max_{f_i}(SM(Q_p, Q_r), VSR(Q_p), CodeBLEU(Q_p, Q_r), TE(f_i))$$

subject to:

$$T_{in}(f_i(Q_o, S)) + T_{out}(Q_p) \leq B, \quad (2)$$

where $T_{in}$ and $T_{out}$ are input and output token counts, and $B$ is the available token budget.

This formalization shows why token optimization must be treated as a multi-objective problem rather than simple prompt shortening.

## IV. Related Work and Research Gap

### A. Prompt Compression and Long-Context Processing

Prompt compression has become an important direction for reducing LLM inference cost and improving long-context usability. LLMLingua introduced prompt compression for accelerated inference by removing less important prompt tokens while maintaining task performance [5]. LongLLMLingua extended this idea to long-context scenarios, where compression is used to improve efficiency and retrieval of relevant information [6]. However, SQL migration differs from general natural language tasks because small token changes may alter syntax, identifiers, control flow, or procedural semantics.

The "lost in the middle" effect demonstrates that LLMs may fail to use relevant information when it appears in the middle of long contexts [4]. This is directly relevant to database migration because schema constraints, function definitions, or critical transformation rules may be buried inside long prompts. Repository-level methods such as RepoCoder use retrieval and generation cycles to improve code completion by selecting relevant context [7]. Yet repository retrieval does not fully solve dialect migration because the problem requires both context selection and semantic-preserving transformation.

Other studies investigate the context of compressed code for issue resolution and model adaptation to compressed inputs [16], [17]. These works confirm the importance of context efficiency, but they do not evaluate Oracle-to-PostgreSQL migration strategies using SQL-specific metrics such as Valid Syntax Rate, Semantic Match, and CodeBLEU.

### B. LLM-Based SQL Generation and Database Migration

Text-to-SQL research has made significant progress in mapping natural language questions to executable SQL queries. Spider established a large-scale cross-domain semantic parsing benchmark [18]. DIN-SQL improved Text-to-SQL performance through decomposed in-context learning and self-correction [8]. PICARD introduced constrained decoding to ensure syntactic validity during SQL generation [9]. Large benchmark studies further showed that LLMs can improve Text-to-SQL generation but still require careful evaluation of execution accuracy, syntax validity, and schema grounding [2].

Recent work on constraint-aware SQL translation shows that explicit constraints can improve SQL generation [3]. LLM-based database migration has also been studied as part of the evolution of knowledge graphs and database systems [1]. However, most SQL-generation studies focus on natural-language-to-SQL tasks rather than source-dialect-to-target-dialect migration. Oracle-to-PostgreSQL migration requires preserving procedural and schema-level semantics, not just generating a valid query from a natural-language prompt.

### C. Code Representation and Structure-Aware Generation

Code representation models such as CodeBERT, GraphCodeBERT, TreeBERT, UniXcoder, and SynCoBERT demonstrate that structural features, syntax trees, data flow, and cross-modal code representations improve code understanding and generation [10]–[14]. StarCoder and CodeGeeX further demonstrate the role of large-scale code pretraining in multilingual code generation [15], [19]. Codex evaluation introduced execution-based assessment for code generation and influenced modern code-generation benchmarks [20].

These studies support the assumption that structure matters in LLM-based code transformation. In Oracle-to-PostgreSQL migration, structural information is encoded in DDL, PL/SQL control flow, schema dependencies, triggers, and stored procedures. Therefore, token optimization strategies that remove or obscure structural information may reduce semantic equivalence even when they reduce token usage.

### D. Code Generation Metrics

Code generation cannot be adequately evaluated using only natural-language similarity metrics. BLEU was originally developed for machine translation [21], but code generation requires structural and semantic evaluation. CodeBLEU improves code evaluation by combining n-gram matching, weighted keyword matching, syntax matching, and data-flow matching [22]. Automated program repair research also emphasizes the importance of execution-based and semantic evaluation for generated code [23].

For SQL migration, exact character-level matching is too strict because equivalent SQL can be written in multiple forms. At the same time, syntax validity alone is insufficient because syntactically valid SQL may not preserve business logic. Therefore, this paper combines VSR, EM, SM, CodeBLEU, and TE.

### E. AI-Driven and Agentic Software Development

Multi-agent software development frameworks such as AutoGen, ChatDev, MetaGPT, and related agentic systems demonstrate that software engineering tasks can be decomposed among agents responsible for planning, generation, testing, review, and coordination [24]–[27]. However, multi-agent workflows can also increase context size because each agent may require role descriptions, task histories, intermediate artefacts, tool outputs, and validation feedback.

Specification-driven tools such as fspec, OpenSpec, spec-driver, MetaSpec, SHOTGUN, Spec Kit, MoAI-ADK, GoopSpec, cc-sdd, Kiro, Spec Kitty, Compose-Lang, and LAP show that specifications are increasingly treated as executable or semi-executable artefacts for AI coding agents. These tools support requirements-to-code workflows, test-driven generation, specification compaction, and agent steering. However, they do not provide a general empirical model for measuring the trade-off between token reduction and migration quality.

### F. Research Gap

The reviewed literature provides strong foundations for prompt compression, long-context processing, SQL generation, code representation, and code generation evaluation. However, the following gap remains:

*Existing studies do not sufficiently address token optimization as a constrained Oracle-to-PostgreSQL migration problem, where cost reduction must be balanced against dialect correctness, procedural logic preservation, identifier semantics, syntactic validity, and semantic equivalence.*

General prompt compression may remove information that is semantically essential for SQL migration. General Text-to-SQL benchmarks do not represent source-dialect-to-target-dialect migration. Code representation models show that structure matters, but they do not define practical token optimization strategies for enterprise SQL migration. Therefore, this paper proposes and evaluates a systematic set of token optimization strategies for LLM-based Oracle-to-PostgreSQL migration.

## V. Proposed Taxonomy of Token Optimization Strategies

### A. Overview of the Proposed Taxonomy

The study evaluates twelve strategies. Each strategy transforms either the input artefact, the system prompt, or both. Table II summarizes the strategies.

TABLE II. Token Optimization Strategies for Oracle-to-PostgreSQL Migration

| No. | Strategy | Technical implementation | Core logic |
|---|---|---|---|
| 1 | Baseline | Unmodified representation | Passes raw Oracle SQL/PL-SQL directly into the LLM context |
| 2 | Context pruning | Comment and prompt stripping | Removes comments and Oracle-specific physical parameters such as TABLESPACE, STORAGE, and PCTFREE |
| 3 | Minification | Whitespace hardening | Collapses redundant whitespace, tabs, and newlines |
| 4 | Semantic compression | DSL mapping | Encodes repetitive SQL structures using compact symbolic forms |
| 5 | Metadata augmentation | Keyword-guided anchor injection | Adds schema properties and key parameters as semantic anchors |
| 6 | Context refactoring | Density adjustment | Removes token-expensive quote characters around identifiers |
| 7 | Schema distillation | Logical core abstraction | Removes secondary procedural definitions and keeps only core schema logic |
| 8 | Adaptive routing | Structural complexity analysis | Selects the optimization method based on input complexity |
| 9 | AST-based minification | Abstract syntax tree parsing | Parses SQL into an AST and serializes it into compact form |
| 10 | Identifier masking | Dynamic token aliasing | Replaces long identifiers with short deterministic aliases |
| 11 | Output constraint enforcement | System prompt guardrails | Restricts formatting, comments, markdown, and conversational output |
| 12 | Hybrid optimization | Dual-bound control | Combines adaptive input compression with output constraints |

The strategies differ in their aggressiveness. Context pruning and minification remove surface-level noise. DSL compression and identifier masking alter representation. Metadata augmentation adds information. Distillation strongly abstracts the code. Adaptive routing attempts to select the most appropriate strategy based on structural complexity. Output-constraint enforcement controls the generated output length rather than the input length.

### B. Base Space Definitions

Building upon the specification-driven development formalization introduced in Section III, we mathematically define the optimization strategies using the following base space variables:

- $Q_o$ – original Oracle SQL/PL-SQL source code;
- $S$ – the specification and context bundle;
- $H$ – the system prompt;
- $G(x, y)$ – the generation function of the LLM, where $x$ is the input prompt and $y$ is the system prompt;
- $Q_p$ – generated PostgreSQL artifact;
- $f_i$ – any token optimization strategy applied to the input, treated as a transformation function.

### C. Baseline

The baseline strategy involves passing the raw, unedited Oracle SQL source code directly into the LLM context buffer without any optimization. It serves as a benchmark to measure baseline LLM performance and token consumption.

$$f_{base}(Q_o) = Q_o, \tag{3}$$

### D. Context Pruning

This method strips out developer comments and Oracle-specific physical infrastructure parameters, such as TABLESPACE, STORAGE, and PCTFREE. Since PostgreSQL manages physical storage allocation differently from Oracle, passing these parameters to the LLM consumes context window space without providing functional value. This strategy may be effective because it removes noise, directly reducing input token usage while preserving the business logic.

The targeted removal of developer comments $C$ and Oracle-specific physical infrastructure parameters $I$ (TABLESPACE, PCTFREE).

$$f_{prune}(Q_o) = Q_o \backslash (C \cup I), \quad (4)$$

*E. Minification*

Minification collapses all redundant whitespaces, tabs, and newline characters within the SQL block. Also, it optimizes the space by removing whitespaces around operators. It may be effective because it maximizes token information density, allowing more actual logic to fit into the same context window limit.

Let $S_{space}$ represent the set of redundant whitespace and newline characters.

$$f_{min}(Q_o) = f_{prune}(Q_o) \setminus S_{space}, \quad (5)$$

*F. Semantic Compression*

This strategy encodes repetitive SQL structures and operators into a compressed, custom Domain Specific Language. For example, it maps long token definitions like CREATE OR REPLACE to shorthand variables like CR:. It may be effective for very large SQL packages.

Let $D$ be the ordered dictionary of substitution pairs, sorted by keyword length in descending order to prevent partial word replacements:

$$D = \{(k_1, v_1), (k_2, v_2), \dots, (k_n, v_n)\}, \quad (6)$$

Since string substitution is a sequential process, it is mathematically represented as a composition of functions. Let $R_{k,v}(x)$ be a function that replaces instances of keyword $k$ with value $v$ within text $x$. The DSL transformation is formalized as the sequential composition of these substitution functions over the minified query:

$$f_{dsl}(Q_o) = (R_{k_n,v_n} \circ R_{k_{n-1},v_{n-1}} \circ \dots \circ R_{k_1,v_1})(f_{min}(Q_o)), \quad (7)$$

*G. Metadata Augmentation*

Rather than removing text, this method extracts schema properties and key parameters and injects them alongside the source code. In this way, the LLM receives additional information, which can improve accuracy and reduce the number of migration errors. Thus, if the quality of the translation improves, the efficiency of token usage will also increase.

Let $K_{meta} \subset M$ be the subset of extracted schema properties or top-utilized keywords $M$. The transformation is mathematically represented as the string concatenation $\oplus$ of the serialized metadata context and the minified query.

$$f_{meta}(Q_o, K_{meta}) = K_{meta} \oplus f_{min}(Q_o), \quad (8)$$

*H. Context Refactoring*

This approach removes unnecessary quote characters from pre-minified code. In database schemas, quotes are often treated as separate tokens by the LLM tokenizer. By identifying and stripping quotes from Oracle identifiers, the total token count is reduced.

Let $I_{quoted}$ represent the set of all standard schema identifiers in $Q$ that are enclosed in quotes, and let $id_{clean}$ be the corresponding identifier with the quotes removed. The transformation sequentially replaces the quoted instances within the minified code.

$$f_{refact}(Q_o) = f_{min}(Q_o)[id \rightarrow id_{clean}], \forall id \in I_{quoted}, \quad (9)$$

*I. Schema Distillation*

Schema distillation removes all secondary procedural code, leaving only the definition of the main schema types and the logical flow. This method can be effective because it retains only the definitions of functions and procedures, significantly reducing the number of input tokens.

$$f_{distil}(Q_o) = Q_{o_{header}}, \quad (10)$$

*J. Adaptive Routing*

Adaptive routing analyzes the structural complexity of the input code and dynamically routes it to the appropriate optimization strategy. If the Oracle code consists of >80% PL/SQL, it applies DSL compression; if it contains 0% PL/SQL, it applies the context refactoring strategy.

A piecewise function dynamically routing $Q_o$ to the most efficient strategy based on the PL/SQL percentage parameter $p \in [0, 100]$ retrieved from the metadata vector $M$.

$$f_{adapt}(Q_o, p) = \begin{cases} f_{dsl}(Q_o) & if\ p > 80 \\ f_{refac}(Q_o) & if\ p = 0 \\ f_{min}(Q_o) & otherwise \end{cases}, \quad (11)$$

*K. AST-Based Minification*

This method parses the raw SQL into an Abstract Syntax Tree using a parser such as sqlglot. Once parsed into a tree, it removes redundant syntax nodes and renders the code back into a minimized string. Unlike regex-based minification, which can accidentally break string literals, AST parsing refers to the grammar of the Oracle dialect.

Let $Parse_{AST}(Q_o)$ represent the function that constructs a directed acyclic graph using the Oracle dialect grammar. Let $Prune_{node}(T)$ be the function that traverses the tree to drop non-essential formatting or redundant nodes, and $Render\ (T)$ be the serialization function that outputs the minimized string:

$$f_{ast}(Q_o) = Render(Prune_{node}(Parse_{AST}(Q_o))), \quad (12)$$

*L. Identifier Masking*

Identifier masking maps verbose table, column, and object names to short variables, e.g., replacing a long table name with X_1. This method minimizes input tokens by replacing these strings.

Let $W$ be the set of all unique word tokens in the minified query $f_{min}(Q_o)$, and let $R$ be the set of reserved SQL and PL/SQL keywords. The set of target identifiers $ID$ is defined as the difference between these sets:

$$ID = W \setminus R, \quad (13)$$

To prevent partial word substitutions, e.g., accidentally replacing the substring "USER" inside "USER_ID", the set $ID$ is ordered by string length in descending order:

$$ID = \{id_1, id_2, ..., id_n\}, \quad (14)$$

A bijective mapping dictionary $Y$ is generated, assigning an incremental short alias to each unique identifier:

$$Y = \{(id_1, X_1), (id_2, X_2), ..., (id_n, X_n)\}, \quad (15)$$

The input query is compressed by sequentially substituting the original identifiers with their corresponding aliases:

$$f_{mask}(Q_o) = f_{min}(Q_o)[id_i \rightarrow X_i], \forall (id_i, X_i) \in Y, \quad (16)$$

Let $Q_{p_{masked}}$ be the raw output generated by the LLM containing the compressed aliases. The final, executable PostgreSQL code $Q_p$ is reconstructed by reversing the substitution process:

$$Q_p = Q_{p_{masked}}[X_i \rightarrow id_i], \forall (id_i, X_i) \in Y, \quad (17)$$

*M. Output Constraint Enforcement*

This method controls the output by injecting explicit system-level instructions that prohibit the LLM from generating newlines, formatting spaces, markdown text, or conversational comments.

Rather than modifying the input $Q_o$, this approach modifies the system prompt $H$ by injecting strict generative constraints $C_{min}$ to prevent the LLM from producing comments or formatting in the output, thereby minimizing $T_{out}$.

$$Q_p = G(Q_o, H \cup C_{min}), \quad (18)$$

*N. Hybrid Optimization*

Hybrid optimization combines adaptive routing with output constraint enforcement. By applying logic to both user and system prompts, it manages token usage on both sides of the migration.

$$Q_p = G(f_{adapt}(Q_o), H \cup C_{min}), \quad (19)$$

## VI. Evaluation Metrics

*A. Valid Syntax Rate*

Valid Syntax Rate measures the percentage of generated PostgreSQL outputs that can be parsed successfully:

$$VSR = \frac{N_{valid}}{N} \times 100\%, \quad (20)$$

where $N_{valid}$ is the number of generated outputs parsed without syntax error, and N is the total number of migration cases.

*B. Exact Match*

Exact Match evaluates whether the generated output is character-identical to the reference PostgreSQL code:

$$EM = \frac{N_{exact}}{N} \times 100\%, \quad (21)$$

where $N_{exact}$ is the number of the match.

Although EM is strict, it is useful for identifying cases where the model exactly reconstructs the expected target.

*C. Semantic Match*

Semantic Match evaluates functional equivalence between the generated PostgreSQL code and the reference PostgreSQL code. Since SQL can be semantically equivalent despite textual differences, SM is based on hybrid AST comparison and expert evaluation if parsing is reliable, as:

$$SM(Q_p, Q_r) = ASTSim(Q_p, Q_r). \quad (22)$$

If parsing is not reliable, we use Subject Matter Experts (SMEs):

$$SM(Q_p, Q_r) = ExpertEval(Q_p, Q_r), \quad (23)$$

This metric captures business logic preservation more effectively than exact matching.

*D. CodeBLEU*

CodeBLEU measures structural similarity using n-gram matching, keyword matching, and code-aware similarity components [22]. In this study, CodeBLEU is used as a structural similarity metric for generated SQL.

$$CodeBLUE = \alpha \cdot BLUE_{ngram} + \beta \cdot Keyword_{Match}, \quad (24)$$

### E. Token Efficiency

Token Efficiency measures semantic return on token cost:

$$TE = \frac{SM}{T_{in}+T_{out}} \times 1000, \quad (25)$$

where $T_{in}$ is the average input token count, and $T_{out}$ is the average output token count. $TE$ is not interpreted alone; it must be analysed together with $SM$ , and $VSR$ because high $TE$ can result from excessive compression and semantic loss.

## VII. Experimental Setup

### A. Dataset Description

The evaluation was conducted using a structured dataset of Oracle SQL and PL/SQL artefacts paired with their corresponding, validated PostgreSQL translations. The dataset is formatted as a parallel code corpus, mapping input_db_query to output_db_query, and encompasses a diverse range of enterprise-level database objects.

The structural complexity of the Oracle queries varies significantly, covering the following key database components:

- procedural logic (PL/SQL);
- dialect-specific exception handling;
- package transformations;
- obfuscated identifiers.

From this broader repository, two specific subsets were drawn for the experiments: a 10-query sample for initial optimization validation and a 100-query sample for robust statistical evaluation of the strategies across varying lengths and complexities.

### B. Experimental Configuration and Model Setup

The experiments evaluate the dialect migration process using GPT-4o as the target LLM. Each Oracle input from the selected datasets was systematically transformed using the twelve token optimization strategies defined in this study. To establish a reliable benchmark for evaluating token savings and potential generation degradation, the baseline strategy passed raw, unoptimized Oracle SQL/PL-SQL code directly into the model context.

### C. Evaluation Protocol and Research Questions

The generated PostgreSQL outputs were evaluated using a multi-dimensional protocol that assesses syntactic correctness, structural fidelity, semantic preservation, and resource costs. Specifically, the outputs were measured using VSR, EM, SM, CodeBLEU, and TE.

The experiment is designed to answer the following research questions:

**RQ1:** Which token optimization strategies reduce token usage without significant semantic degradation?

**RQ2:** Which strategies produce the best trade-off between token efficiency and migration correctness?

**RQ3:** Can aggressive compression improve token efficiency without damaging semantic preservation?

**RQ4:** How do input compression and output control affect syntactic validity and semantic equivalence?

## VIII. Pipeline Architecture

### A. Specification-Driven LLM Migration Pipeline

The proposed pipeline represents Oracle-to-PostgreSQL migration as a specification-driven and context-optimized LLM workflow. The process starts from a developer-defined migration task, where the source Oracle SQL/PL/SQL artefact is interpreted together with business or technical specifications, schema context, and migration rules. Unlike direct code-to-code conversion, the pipeline treats the migration context as a structured artefact that can be selectively optimized before being passed to the LLM. This is important because SDD workflows typically rely on specifications, typed artefacts, code context retrieval, agent orchestration, and validation stages rather than a single prompt-to-code step.

Fig. 1 presents the proposed specification-driven LLM migration pipeline. The pipeline integrates specification artefacts, migration rules, schema context, context optimization methods, LLM-based PostgreSQL generation, and multidimensional evaluation. The figure emphasizes that token optimization is not an isolated preprocessing step, but a controlled transformation layer between the specification-driven development context and the LLM migration engine.

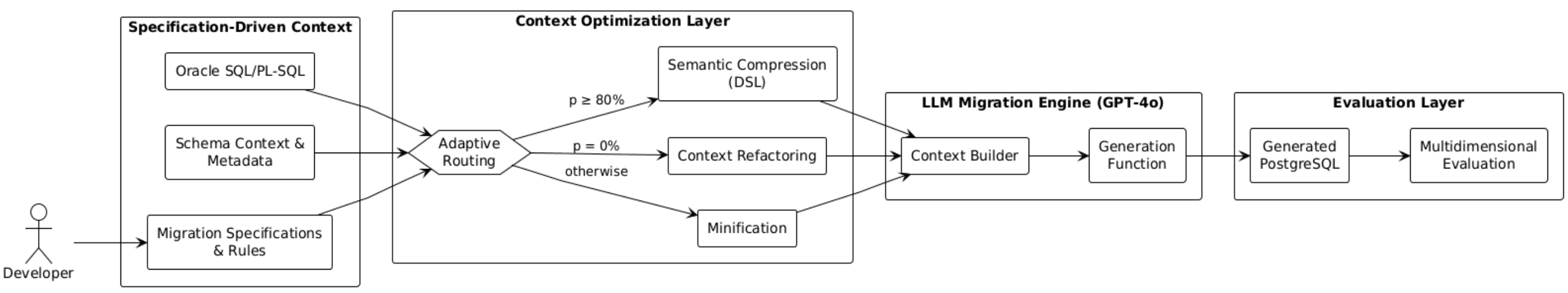

Fig. 1. SDD-Based LLM Migration Pipeline.

As shown in Fig. 1, the developer does not interact only with raw SQL code. The migration input is constructed from specification artefacts, schema context, and Oracle-to-PostgreSQL transformation rules. These artefacts are processed by the context optimization layer. As illustrated, this layer can dynamically route inputs using adaptive routing to apply semantic compression, context refactoring, or minification based on the code's structural complexity. The optimized context is then passed to the LLM generator. The generated PostgreSQL code is evaluated using syntactic, textual, semantic, structural, and efficiency-oriented metrics. This design reflects the study's main assumption: rather than applying a single, uniform token-optimization method, context compression during SQL migration must be performed adaptively based on code complexity.

### B. Token Optimization Trade-Off Model

The second figure explains why token optimization must be treated as a multi-objective problem. The empirical results show that reducing tokens does not automatically improve migration quality. For example, schema distillation strongly reduced token usage and increased Token Efficiency, but it also caused a 44.50 percentage-point decrease in Semantic Match on the 100-query sample. In contrast, adaptive routing produced a more balanced result by reducing input tokens by 8.72% and output tokens by 5.49%, while maintaining 88.40% Semantic Match and increasing Token Efficiency by 6.67%.

Fig. 2 illustrates the trade-off model underlying token optimization in LLM-based database migration. Each optimization strategy may produce positive effects, such as lower input tokens, lower output tokens, lower cost, and higher throughput. However, the same strategy may also introduce negative effects, including semantic drift, syntax invalidity, loss of domain identifiers, or output expansion.

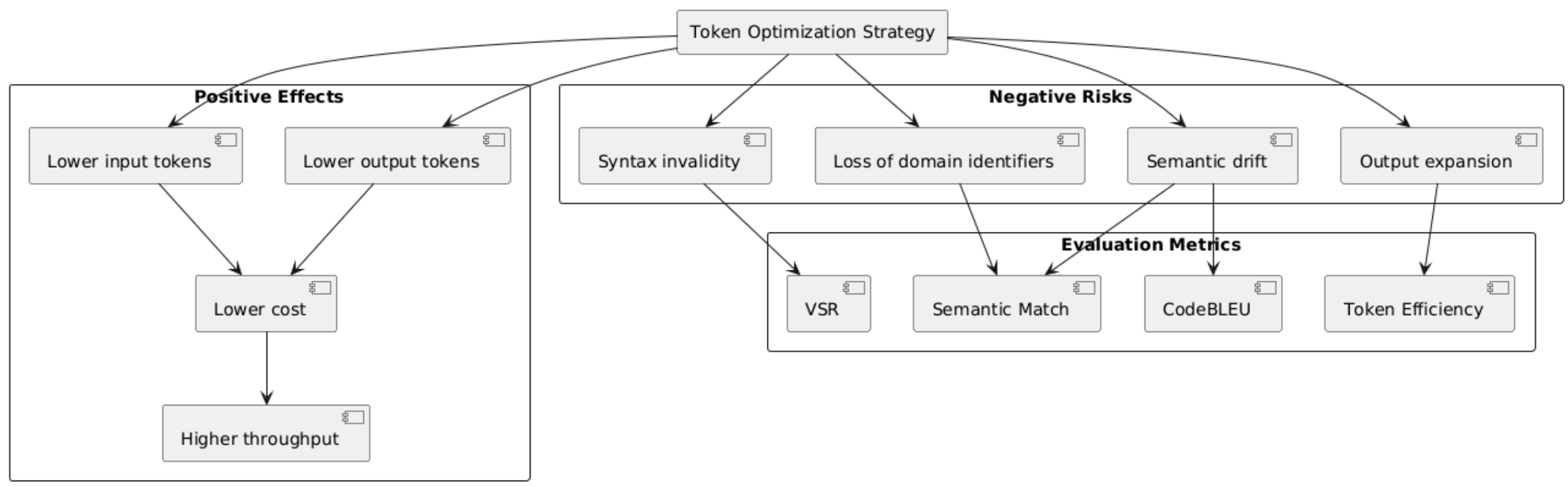

Fig. 2. Token Optimization Trade-Off Model.

Fig. 2 shows that token optimization has a bidirectional effect. On the positive side, fewer input and output tokens reduce computational cost and increase migration throughput. On the negative side, excessive transformation may remove information that is necessary for correct migration. Semantic drift affects Semantic Match, syntax invalidity affects VSR, loss of meaningful identifiers affects semantic interpretation, and output expansion can reduce the expected benefit of input compression. Therefore, Token Efficiency must be interpreted together with VSR, Semantic Match, and CodeBLEU. A high Token Efficiency value alone does not indicate successful migration if semantic equivalence is degraded.

### C. Strategy Selection Pipeline

The third figure converts the empirical findings into a practical decision pipeline. The selection process starts by analysing the structural properties of the Oracle SQL/PL/SQL artefact. The most important features are procedural complexity, the proportion of PL/SQL, the presence of long identifiers, the presence of Oracle-specific physical parameters, and the expected importance of semantic preservation. The experimental results indicate that no single optimization strategy is optimal for all cases. Mild pruning is safest when semantic preservation is critical, while adaptive routing is more suitable for heterogeneous migration batches. Distillation may be acceptable only for simple DDL-oriented cases, but it is unsuitable for procedural logic because it substantially reduces Semantic Match.

Fig. 3 presents the proposed strategy selection algorithm. The algorithm uses structural characteristics of the Oracle artefact and migration objective to select an appropriate token optimization strategy. The purpose of this algorithm is to avoid uniformly applying aggressive compression across all migration cases.

The strategy selection algorithm reflects the empirical behaviour observed in the experiments. If semantic preservation is the main requirement, the baseline or context pruning should be used because pruning preserved Semantic Match almost at the baseline level. If the migration batch is large and structurally heterogeneous, adaptive routing is preferable because it provides the best practical trade-off between token reduction and semantic preservation. If the artefact is DDL-oriented and contains little procedural logic, cautious distillation may be considered, but the result must be verified. For heavy PL/SQL artefacts, distillation and identifier masking should be avoided because they remove or obscure information required for semantic interpretation. After generation, the output must be evaluated using VSR, Semantic Match, CodeBLEU, and Token Efficiency. If the result does not satisfy syntactic and semantic thresholds, the pipeline falls back to a less aggressive optimization strategy.

The proposed migration pipeline consists of four main layers: specification-driven context construction, token optimization, LLM-based PostgreSQL generation, and multidimensional evaluation.

In the first layer, Oracle SQL/PL/SQL artefacts are combined with schema context, migration rules, and specification artefacts.

In the second layer, the input is optimized using pruning, minification, AST-based minification, adaptive routing, output constraints, or other tested strategies.

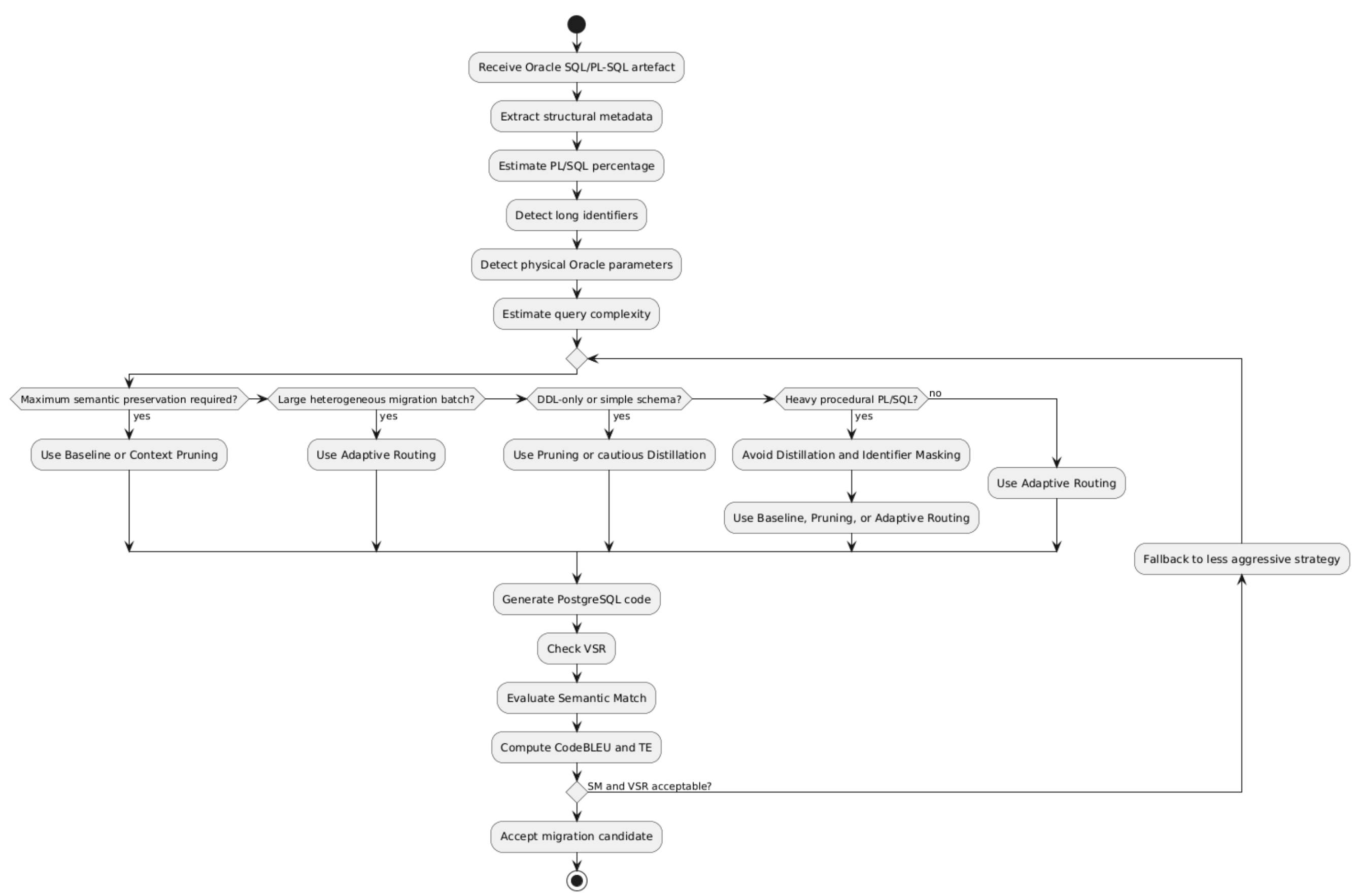

Fig. 3. Strategy Selection Algorithm.

In the third layer, the optimized context is submitted to the LLM generator to produce PostgreSQL code. In the fourth layer, the generated code is evaluated using VSR, EM, SM, CodeBLEU, and TE. The evaluation results determine whether the output is accepted or whether the pipeline falls back to a less aggressive optimization strategy.

## IX. Results

### A. Results on 10 Oracle SQL Queries

Table III presents the results for the 10-query sample.

TABLE III. Results for 10 Oracle SQL Queries

| Strategy | Avg In Tokens | Avg Out Tokens | VSR % | EM % | SM % | CodeBLEU % | TE |
|---|---|---|---|---|---|---|---|
| Original | 431.8 | 442.5 | 80.0 | 0.0 | 80.5 | 79.67 | 0.921 |
| Pruning | 421.4 | 422.1 | 80.0 | 10.0 | 83.0 | 86.20 | 0.984 |
| Minification | 388.4 | 426.3 | 80.0 | 0.0 | 74.0 | 80.83 | 0.908 |
| DSL | 385.8 | 469.0 | 80.0 | 0.0 | 76.5 | 80.09 | 0.895 |
| Metadata | 400.4 | 455.4 | 90.0 | 10.0 | 78.5 | 82.80 | 0.917 |
| Adaptive | 392.9 | 440.9 | 80.0 | 0.0 | 77.5 | 81.36 | 0.929 |
| Context Refactoring | 392.4 | 439.9 | 70.0 | 0.0 | 78.5 | 80.36 | 0.943 |
| Distillation | 43.4 | 66.6 | 70.0 | 0.0 | 34.5 | 59.17 | 3.136 |
| AST-Based Minification | 392.3 | 428.1 | 80.0 | 10.0 | 74.0 | 84.12 | 0.902 |
| Identifier Masking | 288.4 | 442.0 | 80.0 | 10.0 | 67.0 | 79.82 | 0.917 |
| Prompt Restricted | 392.4 | 379.3 | 80.0 | 0.0 | 74.0 | 83.83 | 0.959 |
| Hybrid Optimization | 392.9 | 380.7 | 80.0 | 0.0 | 65.5 | 84.09 | 0.847 |

The 10-query sample shows that pruning improves Semantic Match from 80.5% to 83.0% and increases CodeBLEU from 79.67% to 86.20%. This indicates that removing non-functional noise can improve generation quality. However, aggressive distillation produces a high TE value of 3.136 while reducing SM to 34.5%, demonstrating that token efficiency alone can be misleading.

### B. Results on 100 Oracle SQL Queries

Table IV presents the results for the 100-query sample.

TABLE IV. Results for 100 Oracle SQL Queries

| Strategy | Avg In Tokens | Avg Out Tokens | VSR % | EM % | SM % | CodeBLEU % | TE |
|---|---|---|---|---|---|---|---|
| Original | 496.90 | 495.57 | 71.0 | 5.0 | 89.80 | 85.48 | 0.90 |
| Pruning | 490.99 | 490.24 | 70.0 | 5.0 | 89.75 | 86.07 | 0.91 |
| Minification | 450.10 | 480.70 | 73.0 | 4.0 | 82.57 | 87.36 | 0.8871 |
| DSL | 448.40 | 499.00 | 64.0 | 6.0 | 83.15 | 87.46 | 0.8777 |
| Metadata | 462.00 | 667.70 | 81.0 | 6.0 | 80.67 | 86.08 | 0.7141 |
| Adaptive | 453.56 | 468.35 | 71.0 | 4.0 | 88.40 | 84.94 | 0.96 |
| Context Refactoring | 453.33 | 632.81 | 74.0 | 5.0 | 88.40 | 85.41 | 0.81 |
| Distillation | 90.82 | 125.70 | 77.0 | 2.0 | 45.30 | 66.56 | 2.09 |
| AST-Based Minification | 453.60 | 636.80 | 74.0 | 5.0 | 82.10 | 87.65 | 0.7530 |
| Identifier Masking | 331.44 | 380.62 | 62.0 | 3.0 | 59.72 | 78.43 | 0.8387 |
| Prompt Restricted | 496.90 | 420.99 | 64.0 | 1.0 | 79.02 | 86.62 | 0.8609 |
| Hybrid Optimization | 392.9 | 380.7 | 80.0 | 0.0 | 65.5 | 84.09 | 0.847 |

The 100-query experiment provides a more stable picture. The baseline achieves 89.80% Semantic Match. Pruning preserves almost the same value, 89.75%, while slightly improving CodeBLEU from 85.48% to 86.07%. Adaptive routing achieves the best practical trade-off, maintaining 88.40% Semantic Match and improving TE to 0.96.

### C. *Difference from Baseline*

Table V shows the difference between each strategy and the baseline on the 100-query sample.

TABLE V. DIFFERENCE FROM BASELINE ON 100 ORACLE SQL QUERIES

| Strategy | Avg In Tokens | Avg Out Tokens | VSR | EM | SM | CodeBLEU | TE |
|---|---|---|---|---|---|---|---|
| Pruning | -1.19% | -1.08% | -1.00 pp | 0.00 pp | -0.05 pp | +0.59 pp | +1.11% |
| Minification | -9.42% | -3.00% | +2.00 pp | -1.00 pp | -7.23 pp | +1.88 pp | -1.43% |
| DSL | -9.76% | +0.69% | -7.00 pp | +1.00 pp | -6.65 pp | +1.98 pp | -2.48% |
| Metadata | -7.02% | +34.73% | +10.00 pp | +1.00 pp | -9.13 pp | +0.60 pp | -20.66% |
| Adaptive | -8.72% | -5.49% | 0.00 pp | -1.00 pp | -1.40 pp | -0.54 pp | +6.67% |
| Context Refactoring | -8.77% | +27.69% | +3.00 pp | 0.00 pp | -1.40 pp | -0.07 pp | -10.00% |
| Distillation | -81.72% | -74.64% | +6.00 pp | -3.00 pp | -44.50 pp | -18.92 pp | +132.22 % |
| AST-Based Minification | -8.71% | +28.50% | +3.00 pp | 0.00 pp | -7.70 pp | +2.17 pp | -16.33% |
| Identifier Masking | -33.30% | -23.20% | -9.00 pp | -2.00 pp | -30.08 pp | -7.05 pp | -6.81% |
| Prompt Restricted | 0.00% | -15.05% | -7.00 pp | -4.00 pp | -10.78 pp | +1.14 pp | -4.34% |
| Hybrid Optimization | -8.72% | -12.28% | -2.00 pp | -3.00 pp | -12.23 pp | +0.08 pp | -2.97% |

## X. DISCUSSION

### A. *Mild Pruning as a Safe Optimization Strategy*

Pruning provides the safest optimization profile. It reduces input tokens by 1.19% and output tokens by 1.08%, while Semantic Match decreases by only 0.05 percentage points. *CodeBLEU* increases by 0.59 percentage points, and *TE* improves by 1.11%. This suggests that Oracle-specific physical storage parameters and comments often consume tokens without contributing significantly to migration semantics.

### B. *Adaptive Routing as a Practical Production Strategy*

Adaptive routing provides the best practical trade-off among the tested strategies. It reduces input tokens by 8.72% and output tokens by 5.49%, while Semantic Match decreases by only 1.40 percentage points. *TE* increases by 6.67%. These results indicate that dynamic selection of optimization intensity based on structural complexity is more effective than applying a uniform compression strategy to all queries.

### C. *Aggressive Compression Causes Semantic Drift*

Distillation dramatically reduces the number of input and output tokens but causes severe semantic degradation. It reduces input tokens by 81.72% and output tokens by 74.64%, but Semantic Match decreases by 44.50 percentage points. Although *TE* increases by 132.22%, this value reflects a mathematical ratio rather than practical migration success. The result shows that *TE* must not be interpreted independently of Semantic Match.

### D. *Syntactic Validity Does Not Guarantee Semantic Equivalence*

Metadata augmentation achieves the highest *VSR* of 81.0%, improving syntax validity by 10 percentage points compared with the baseline. However, Semantic Match decreases by 9.13 percentage points. This indicates that additional metadata can help the model produce grammatically valid PostgreSQL but may also introduce semantic distortion or hallucinated assumptions. Therefore, *VSR* must be combined with *SM* and *CodeBLEU*.

### E. *Identifier Masking Is Risky for Database Migration*

Identifier masking reduces input tokens by 33.30% and output tokens by 23.20%, but it reduces Semantic Match by 30.08 percentage points. This result confirms that database identifiers carry domain semantics. Table names, column names, procedure names, and variable names often help the model infer business logic. Replacing them with generic aliases may reduce token cost but damage comprehension.

### F. *Output Constraint Enforcement Reduces Output Tokens but Harms Quality*

Prompt-restricted output reduces output tokens by 15.05%, but *VSR* decreases by 7 percentage points and Semantic Match decreases by 10.78 percentage points. This suggests that overly strict output constraints can interfere with the model's ability to generate correct SQL. Output control should therefore be applied cautiously.

### G. *Hybrid Optimization Does Not Guarantee Synergy*

Hybrid optimization combines adaptive routing with output constraints. Although it reduces input tokens by 8.72% and output tokens by 12.28%, Semantic Match decreases by 12.23 percentage points. This shows that combining multiple optimization techniques does not necessarily produce a synergistic effect. Instead, errors and representation losses can accumulate.

## XI. PROPOSED STRATEGY SELECTION MODEL

Based on the results, the following decision model is proposed:

1. If maximum semantic preservation is required, use **Baseline** or **Context Pruning**.
2. If scalable migration is required with moderate token savings, use **Adaptive Routing**.
3. If the input is simple DDL with low procedural complexity, **Distillation** may be used cautiously.
4. If identifiers carry domain semantics, avoid **Identifier Masking**.
5. If output cost is the main constraint, use **Prompt Restricted** only after verifying VSR and SM.
6. **Avoid applying** aggressive compression uniformly across heterogeneous SQL/PL-SQL artefacts.
7. **Always combine** token metrics with semantic and syntactic metrics.

## Conclusions

This paper investigated token optimization strategies for LLM-based Oracle-to-PostgreSQL migration. The study formalized token optimization as a constrained transformation problem inside specification-driven development and evaluated twelve strategies using *VSR*, *EM*, *SM*, *CodeBLEU*, and *TE*.

The results show that token optimization must not be reduced to prompt shortening. Mild pruning is the safest strategy, preserving Semantic Match almost at the baseline level. Adaptive routing provides the best practical balance between token reduction and semantic preservation, reducing input tokens by 8.72% and output tokens by 5.49% while maintaining 88.40% Semantic Match and improving Token Efficiency by 6.67%. In contrast, aggressive strategies such as schema distillation and identifier masking produce strong token reductions but cause substantial semantic degradation.

The main conclusion is that successful LLM-based database migration relies on adaptive context compression rather than uniform token reduction. By dynamically routing inputs based on their structural complexity, the proposed pipeline achieves an optimal balance between token efficiency and migration correctness. Our multidimensional evaluation confirms that while token cost, syntactic validity, and semantic equivalence must be evaluated together, it is the adaptive strategy selection that prevents semantic drift.

Future work should extend the dataset, evaluate additional LLMs, integrate execution-based PostgreSQL testing, and improve adaptive strategy selection based on static analysis of Oracle SQL/PL-SQL complexity.

## References

[1] S. Zhao, Q. Zhang, and M. Lan, "LLM-powered database migration: A framework for knowledge graph system evolution," Alexandria Engineering Journal, vol. 130, pp. 198–207, 2025, doi: 10.1016/j.aej.2025.08.014.

[2] D. Gao, H. Wang, Y. Li, X. Sun, Y. Qian, B. Ding, and J. Zhou, "Text-to-SQL empowered by large language models: A benchmark evaluation," Proceedings of the VLDB Endowment, vol. 17, no. 5, pp. 1132–1145, 2024, doi: 10.14778/3641204.3641221.

[3] T. Ren, C. Ke, Y. Fan, Y. Jing, Z. He, K. Zhang, and X. S. Wang, "The power of constraints in natural language to SQL translation," Proceedings of the VLDB Endowment, vol. 18, no. 7, pp. 2097–2111, 2025, doi: 10.14778/3734839.3734847.

[4] N. F. Liu et al., "Lost in the middle: How language models use long contexts," Transactions of the Association for Computational Linguistics, vol. 12, pp. 157–173, 2024, doi: 10.1162/tacl_a_00638.

[5] H. Jiang et al., "LLMLingua: Compressing prompts for accelerated inference of large language models," in Proc. Conf. Empirical Methods in Natural Language Processing (EMNLP), 2023, pp. 13358–13376, doi: 10.18653/v1/2023.emnlp-main.825.

[6] H. Jiang et al., "LongLLMLingua: Accelerating and enhancing LLMs in long context scenarios via prompt compression," in Proc. Annual Meeting of the Association for Computational Linguistics (ACL), 2024, doi: 10.48550/arXiv.2310.06839.

[7] F. Zhang et al., "RepoCoder: Repository-level code completion through iterative retrieval and generation," in Proc. Conf. Empirical Methods in Natural Language Processing (EMNLP), 2023, doi: 10.48550/arXiv.2303.12570.

[8] M. Pourreza and D. Rafiei, "DIN-SQL: Decomposed in-context learning of text-to-SQL with self-correction," in Proc. Advances in Neural Information Processing Systems (NeurIPS), 2023, doi: 10.48550/arXiv.2304.11015.

[9] T. Scholak, N. Schucher, and D. Bahdanau, "PICARD: Parsing incrementally for constrained auto-regressive decoding from language models," in Proc. Conf. Empirical Methods in Natural Language Processing (EMNLP), 2021, pp. 9895–9901, doi: 10.18653/v1/2021.emnlp-main.779.

[10] Z. Feng et al., "CodeBERT: A pre-trained model for programming and natural languages," in Findings of the Association for Computational Linguistics: EMNLP, 2020, pp. 1536–1547, doi: 10.18653/v1/2020.findings-emnlp.139.

[11] D. Guo et al., "GraphCodeBERT: Pre-training code representations with data flow," in Proc. International Conference on Learning Representations (ICLR), 2021.

[12] X. Jiang et al., "TreeBERT: A tree-based pre-trained model for programming language," in Proc. Conference on Uncertainty in Artificial Intelligence (UAI), 2021, doi: 10.48550/arXiv.2105.12485.

[13] D. Guo et al., "UniXcoder: Unified cross-modal pre-training for code representation," in Proc. Annual Meeting of the Association for Computational Linguistics (ACL), 2022, pp. 7212–7225, doi: 10.18653/v1/2022.acl-long.499.

[14] X. Wang et al., "SynCoBERT: Syntax-guided multi-modal contrastive pre-training for code representation," arXiv:2108.04556, 2021.

[15] R. Li et al., "StarCoder: May the source be with you!" arXiv:2305.06161, 2023.

[16] A. Chevalier et al., "Adapting language models to compress contexts," arXiv:2305.14788, 2023.

[17] H. Jia et al., "Compressing code context for LLM-based issue resolution," arXiv:2603.28119, 2026.

[18] T. Yu et al., "Spider: A large-scale human-labeled dataset for complex and cross-domain semantic parsing and text-to-SQL task," in Proc. Conf. Empirical Methods in Natural Language Processing (EMNLP), 2018, pp. 3911–3921, doi: 10.18653/v1/D18-1425.

[19] Q. Zheng et al., "CodeGeeX: A pre-trained model for code generation with multilingual evaluations," in Proc. ACM SIGKDD International Conference on Knowledge Discovery and Data Mining (KDD), 2023, doi: 10.48550/arXiv.2303.17568.

[20] M. Chen et al., "Evaluating large language models trained on code," arXiv:2107.03374, 2021.

[21] K. Papineni, S. Roukos, T. Ward, and W. Zhu, "BLEU: A method for automatic evaluation of machine translation," in Proc. Annual Meeting of the Association for Computational Linguistics (ACL), 2002, pp. 311–318, doi: 10.3115/1073083.1073135.

[22] S. Ren et al., "CodeBLEU: A method for automatic evaluation of code synthesis," arXiv:2009.10297, 2020.

[23] C. Xia, Y. Wei, and L. Zhang, "Automated program repair in the era of large pre-trained language models," in Proc. IEEE/ACM International Conference on Software Engineering (ICSE), 2023, pp. 1482–1494, doi: 10.1109/ICSE48619.2023.00129.

[24] Q. Wu et al., "AutoGen: Enabling next-gen LLM applications via multi-agent conversation framework," arXiv:2308.08155, 2023.

[25] C. Qian et al., "ChatDev: Communicative agents for software development," in Proc. Annual Meeting of the Association for Computational Linguistics (ACL), 2024.

[26] S. Hong et al., "MetaGPT: Meta programming for a multi-agent collaborative framework," in Proc. International Conference on Learning Representations (ICLR), 2024.

[27] J. Li et al., "Agent-oriented software engineering with large language models," arXiv:2402.01680, 2024.

[28] M. de la Parte, Y. Wang, J.-F. Martínez-Ortega, and N. L. Martínez, "A secure and efficient LLM-based system for natural language query-to-SQL translation in IoT data management," IEEE Access, 2026, doi: 10.1109/ACCESS.2026.3665980.

[29] V. Y. Lyashkevych, M. Y. Lyashkevych, and R. Y. Shuvar, "Definition and formalization of the software functional state concept throughout the development life cycle," Electronics and Information Technologies, vol. 32, pp. 151–170, 2025, doi: 10.30970/eli.32.11.

[30] O. Grynets, V. Lyashkevych, A. Dolichnyi, R. Piznak, T. Zelenyy, and V. Morozov, "Specification-based code–text–code reengineering for LLM-mediated software evolution," arXiv:2605.25232, 2026.

[31] O. Grynets and V. Lyashkevych, "Unified architecture metamodel of information systems developed by generative AI," arXiv:2604.00171, 2026.

[32] V. Lyashkevych, "Evolution-aware specification-driven synthesis of intelligent monitoring systems by large language models," 2026.

[33] V. Lyashkevych, "Intelligent monitoring as an information technology for context-aware decision-making strategy selection," 2026.

[34] V. Kutsan and V. Lyashkevych, "Semantic Kernel usage for orchestration of multi-agent LLM-based systems to solve the tasks which require dynamic involvement of new agents," 2026.

[35] E. Surk, G. G. Menekse Dalveren, and M. Derawi, "Reducing information asymmetry in software product management: An LLM-based reverse engineering framework," Applied Sciences, vol. 16, no. 6, Art. no. 2801, 2026, doi: 10.3390/app16062801.

[36] V. Y. Lyashkevych, "Sustainable optimization of consolidated data processing algorithms based on machine learning and genetic algorithms," Electronics and Information Technologies, vol. 32, pp. 39–54, 2025, doi: 10.30970/eli.32.3.

[37] M. Lyashkevych, V. Lyashkevych, and R. Shuvar, "Risks' attribute values evaluation in software engineering by Monte Carlo simulation," in Proc. IEEE 13th International Conference on Electronics and Information Technologies (ELIT), 2023, pp. 137–141, doi: 10.1109/ELIT61488.2023.10310775.

[38] NIST, "Secure Software Development Framework (SSDF), Version 1.1: Recommendations for mitigating the risk of software vulnerabilities," NIST Special Publication 800-218, 2022, doi: 10.6028/NIST.SP.800-218.